\documentclass{ws-ijmpe}
\usepackage[super,compress]{cite}
\begin{document}

\markboth{R.L. Xu, C. Wu, and Z. Z. Ren}{Superfluidity of $\Lambda$
hyperons in warm strange hadronic star matter}

\catchline{}{}{}{}{}

\title{Superfluidity of $\Lambda$ hyperons in warm strange hadronic star matter}

\author{\footnotesize Renli Xu}

\address{Key Laboratory of Modern Acoustics and Department of
Physics,\\ Nanjing University, Nanjing 210093, P. R. China\\
xurenli.phy@gmail.com}

\author{Chen Wu}
\address{Shanghai Institute of Applied Physics, Chinese Academy of Sciences,\\ Shanghai 201800, P. R. China\\
wuchenoffd@gmail.com}

\author{Zhongzhou Ren}
\address{Key Laboratory of Modern Acoustics and Department of
Physics, \\Nanjing University, Nanjing 210093, P. R. China\\
Center of Theoretical Nuclear Physics, National Laboratory of
Heavy-Ion Accelerator, \\Lanzhou 730000, P. R. China\\ Joint Center of
Nuclear Science and Technology, \\Nanjing University, Nanjing
210093, P. R. China\\zren@nju.edu.cn}

\maketitle

\begin{history}
\received{15 May 2014}
\accepted{22 October 2014}
\end{history}

\begin{abstract}
In this work we evaluate the $^1S_0$ superfluidity of $\Lambda$
hyperons in $\beta$-stable strange hadronic matter. We investigate
the equation of state (EOS) of hadronic matter at finite values of baryon
density and temperature in the relativistic mean field (RMF) theory.
Effects of the introduced isoscalar-isovector cross-interaction term
on the $\Lambda$ superfluidity are investigated systematically. In
addition, the temperature effects on the superfluidity of $\Lambda$
hyperons in hadronic matter are discussed. It is found that the density region and
the magnitude of the $\Lambda$ pairing gap are
dependent on the cross-interaction term. The obtained maximal
critical temperature of $\Lambda$ superfluid is around $10^9$ K.
\end{abstract}


\ccode{PACS number(s): 26.60.-c, 21.30.Fe, 26.60.Kp}


\section{Introduction}

The nuclear physics of hadronic matter has become a hot topic that
connects astrophysics with extreme high-density nuclear physics
\cite{Schaffner1,Muller,Gal,Trumper,Akmal,Schaffner2,Typel,Klahn,Lavagno,Panda1,Shen}.
With the progress of astronomical observation and nuclear
experiment, astrophysics phenomena and nuclear physics are combined
more and more tightly. Meantime, neutron star matters have also been
attracting much interest because they offer a good chance of
studying the occurrence of superfluidity in nuclear matter
\cite{Kucharek,Alm,Sandulescu,Baldo,Aguirre,Bombaci,Zhou}.
Superfluidity of baryons in hadronic star matters is expected to
have a number of consequences directly related to observational
effects, such as cooling rates and the glitches in rotational rates
that are observed in a number of pulsars \cite{Pethick,Pines,Alpar}.
It is well believed that baryon superfluidity plays an incisive role
in the thermal evolution of neutron stars
\cite{Page,Heiselberg,Dean}. Baryon pairing may significantly
suppress cooling rates that rely on neutrino emission from the
direct Urca process \cite{Tsuruta,Takatsuka,Balberg}.

In the neutron star interior, hyperons are possible to appear
through the weak interaction with the fast rise of the baryon
density. Generally, $\Lambda$ particle, the lightest
hyperon with an attractive potential, is the first hyperon appearing in nuclear matter
\cite{Wang,Glendenning1,Panda}. The $^1S_0$ $\Lambda$ superfluid may
occur in nuclear medium due to the attractive $\Lambda\Lambda$
interaction \cite{Takatsuka1,Tanigawa,Takatsuka2,Takatsuka3}. Over
the last decade, there have been several literatures about the
nucleon and hyperon superfluidity in hadronic matter. In the work of
Alm {\it et al.} \cite{Alm}, the superfluid $^3D_2$ proton-neutron
pairing in dense isospin-asymmetric nuclear matter is investigated
in terms of the real-time Green functions approach. They found that
the critical temperature associated with the transition to the
superfluid phase becomes strongly suppressed with increasing isospin
asymmetry. In Ref. \refcite{Balberg}, Balberg and Barnea studied the
$^1S_0$ gap energies of $\Lambda$ hyperons in neutron star matter,
using the $G$-matrix effective interaction. They found that a gap
energy of several tenths of a MeV is expected for a $\Lambda$ Fermi
momenta, $k_F(\Lambda)$, below 1.3 fm$^{-1}$. In Refs.
\citen{Takatsuka1,Takatsuka2} and \citen{Takatsuka3}, Takatsuka and
Tamagaki investigated superfluidity of $\Lambda$ hyperons by a
realistic approach using bare $\Lambda\Lambda$ interactions and the
effective mass of $\Lambda$ based on the G-matrix calculations.
Their calculation predicts $\Lambda$ superfluid can exist in a
density region between 2$\rho_0$ and (3$-$4.5)$\rho_0$
\cite{Takatsuka1}, depending on hyperon core models. Besides, the
predicted critical temperature of $\Lambda$ superfluidity is around
$10^9$ K in hyperon-mixed neutron star cores \cite{Takatsuka2}.
Tanigawa {\it et al.} have investigated the $\Lambda\Lambda$ pairing
in binary mixed matter of nucleons and $\Lambda$ hyperons, using the
relativistic Hartree-Bogoliubov model combined with the relativistic
mean field (RMF) interaction in Ref. \citen{Tanigawa}. Therein, it
is found that the value of the $\Lambda\Lambda$ pairing gap
decreases as the background nucleon density increases. In the work
of Wang and Shen \cite{Wang}, the $^1S_0$ $\Lambda$ superfluidity in
neutron star matter and neutron stars has been investigated by
employing several $\Lambda\Lambda$ interactions based on the
Nijmegen models. It is found that the maximal pairing gap obtained
is a few tenths of a MeV, and the magnitude and the density region
of the pairing gap are dependent on the $\Lambda\Lambda$
interaction.

In this work, we focus on the temperature effects on the $^1S_0$
superfluidity of $\Lambda$ hyperons in strange hadronic matter by
means of the RMF model with the inclusion of the full octet of
baryons. The RMF is a pioneering framework to describe the nuclear
system as a relativistic many-body system of baryons and mesons,
which has been widely used to investigate the properties of finite
nuclei and nuclear matter
\cite{Ring1,Ring2,Lalazissis1,Lalazissis2,Lalazissis3,Todd1,Todd2,Fattoyev,Agrawal,Ren1,Ren2,Wu}.
Along this direction, many important extensions of RMF theory have
been made, for example, the additional isoscalar-isovector
cross-interaction term was introduced into the extended RMF model
recently \cite{Todd1,Todd2,Fattoyev}. The additional
isoscalar-isovector coupling term is proved to play an important
role in softening the symmetry energy at high densities and reducing
the neutron skin thickness in heavy nuclei. In this work, we will
systematically investigate the influence of the isoscalar-isovector
coupling term on the superfluidity of $\Lambda$ hyperons.
Furthermore, we investigate the temperature effects on the $\Lambda$
superfluidity in nuclear medium in terms of RMF model for the first
time. In our calculation, we firstly investigate the equation of
state (EOS) of strange hadronic matter at finite values of baryon
density and temperature, then solve the finite-temperature gap
equation to discuss the $\Lambda$ superfluid in hadronic matter.

The organization of this paper is as follows. In Sec. 2, we
outline the theoretical framework of the RMF theory for the hadronic
matter at finite temperature. In Sec. 3, we briefly describe the
energy gap equation for $\Lambda$ hyperon pairing. The model
parameters are discussed in Sec. 4. In Sec. 5, numerical
results and discussions are presented. Finally, the main conclusions
are summarized in Sec. 6.

\section{Formulas of the RMF model}

In the RMF theory, the nuclear interaction is usually described by
the exchange of three mesons: the isoscalar meson $\sigma$ which
produces the medium range attraction, the isoscalar-vector meson
$\omega$ responsible for the short range repulsion, and the
isovector-vector meson $\rho$ reproducing the correct value of the
empirical symmetry energy \cite{Serot}. Here, the cross-interaction
term $\omega^2\rho^2$ is included, which was introduced to soften
the symmetry energy at high densities \cite{Todd1}. The baryons
considered in this work include the full octet of the lightest ones
($N$($\emph{p}$, $\emph{n}$), $\Lambda$, $\Sigma^+$, $\Sigma^0$,
$\Sigma^-$, $\Xi^0$, $\Xi^-$) originally investigated by Glendenning
\cite{Glendenning}. The total Lagrangian density $\mathcal {L}$
takes the form
\begin{eqnarray}
\mathcal{L}&=&\sum_j\bar{\psi}_j[i\gamma^\mu\partial_\mu-M_j+g_{\sigma
j}\sigma- g_{\omega j}\gamma^\mu\omega_\mu-\frac{g_{\rho
j}}{2}\gamma^\mu\tau\cdot\rho_\mu
]\psi_j+\frac{1}{2}\partial^\mu\sigma\partial_\mu\sigma\nonumber\\&&-\frac{1}{2}m^2_\sigma
\sigma^2-\frac{1}{4}\Omega^{\mu\nu}\Omega_{\mu\nu}+\frac{1}{2}m^2_\omega\omega^\mu\omega_\mu
-\frac{1}{4}G^{\mu\nu}G_{\mu\nu}+\frac{1}{2}m^2_\rho\rho^\mu\rho_\mu\nonumber\\&&-U_{eff}(\sigma,
\omega^\mu,
\rho^\mu)+\sum_l\bar{\psi}_l[i\gamma^\mu\partial_\mu-m_l]\psi_l,
\end{eqnarray}
where the index $j$ runs over the full octet of baryons, and $l$
represents electrons and muons ($e$ and $\mu$). $M_j$ denotes the
vaccum baryon mass of index $j$. The antisymmetric tensors of the
vector mesons are taken as the usual forms
$\Omega_{\mu\nu}=\partial_\mu\omega_\nu-\partial_\nu\omega_\mu$ and
$G_{\mu\nu}=\partial_\mu\rho_\nu-\partial_\nu\rho_\mu$. $g_{\sigma
j}$, $g_{\omega j}$ and $g_{\rho j}$ are the coupling constants
between the baryon and $\sigma$ meson, $\omega$ meson, and $\rho$
meson, respectively. The nonlinear self-interacting terms of
$\sigma$, $\omega$ and the isoscalar-isovector cross-interaction are
taken as
\begin{eqnarray}
U_{eff}(\sigma,\omega^\mu,\rho^\mu)&=&
\frac{\kappa}{3!}(g_{\sigma N}\sigma)^3+\frac{\lambda}{4!}(g_{\sigma N}\sigma)^4-\frac{\zeta}{4!}(g^2_{\omega N}\omega_\mu\omega^\mu)^2\nonumber\\
&&-\Lambda_\mathrm{v}(g^2_{\rho N}\rho_\mu\rho^\mu)(g^2_{\omega
N}\omega_\mu\omega^\mu).
\end{eqnarray}
By virtue of translational and rotational invariance, the meson
fields are constant in infinite nuclear matter. As a consequence,
the field equations under the mean-field approximation have the
following form
\begin{eqnarray}
(i\gamma^{\mu}\partial_\mu-M^*_j-g_{\omega j}\gamma^0\omega-\frac{g_{\rho j}}{2}\gamma^0\tau_{3j}\rho)\psi_j=0,\\
m^2_\sigma\sigma+\frac{\kappa}{2}g^3_{\sigma N}\sigma^2+\frac{\lambda}{6}g^4_{\sigma N}\sigma^3=\sum_ig_{\sigma i}\rho^S_i,\\
m^2_\omega\omega+\frac{\zeta}{6}g^4_{\omega
N}\omega^3+2\Lambda_\mathrm{v}g^2_{\rho N} g^2_{\omega N}\rho^2\omega=\sum_ig_{\omega i}\rho^B_i,\\
m^2_\rho\rho+2\Lambda_\mathrm{v}g^2_{\rho N} g^2_{\omega
N}\omega^2\rho=\sum_i\frac{g_{\rho i}}{2}\tau_{3i}\rho^B_i.
\end{eqnarray}
The effective mass of baryon octet, in equation (3), is given as
\begin{eqnarray}
M^*_j=M_j-g_{\sigma j}\sigma.
\end{eqnarray}
$\rho^S_i$ and $\rho^B_i$ are the baryon scalar density and the
baryon density of the particle symbolled by $i$, respectively. They
are written as
\begin{eqnarray}
\rho^S_i=\frac{2}{(2\pi)^3}\int d^3k \frac{M^*_i}{\sqrt{k^2+{M^*_i}^2}}[f_i(k)+\overline{f}_i(k)],\\
\rho^B_i=\frac{2}{(2\pi)^3}\int d^3k [f_i(k)-\overline{f}_i(k)],
\end{eqnarray}
where $f_i(k)$ and $\overline{f}_i(k)$ are the fermion particle
distribution and antiparticle distribution:
\begin{eqnarray}
f_i(k)=\frac{1}{\mathrm{exp}\{(\sqrt{k^2+{M^*_i}^2}-\nu_i)/T\}+1},\\
\overline{f}_i(k)=\frac{1}{\mathrm{exp}\{(\sqrt{k^2+{M^*_i}^2}+\nu_i)/T\}+1},
\end{eqnarray}
with $\nu_i$ being the effective chemical potential, related to the
chemical potential $\mu_i$ as
\begin{eqnarray}
\nu_i=\mu_i-g_{\omega i}\omega-g_{\rho i}\tau_{3i}\rho.
\end{eqnarray}
For the hadronic matter with baryons and charged leptons, the
$\beta$-equilibrium conditions under the weak processes are given by
\begin{eqnarray}
\mu_p=\mu_{\Sigma^+}=\mu_n-\mu_e,\\
\mu_\Lambda=\mu_{\Sigma^0}=\mu_{\Xi^0}=\mu_n,\\
\mu_{\Sigma^-}=\mu_{\Xi^-}=\mu_n+\mu_e,\\
\mu_\mu=\mu_e,
\end{eqnarray}
and the charge neutrality condition is fulfilled by
\begin{eqnarray}
\rho_p+\rho_{\Sigma^+}=\rho_{\Sigma^-}+\rho_{\Xi^-}+\rho_e+\rho_\mu,
\end{eqnarray}
where $\rho_i$ is the number density of particle $i$. At a given
baryon density $\rho_B$ and a given temperature $T$, the Dirac
equation (3) can be solved exactly with plane waves as solutions,
while the coupled eqs. (4)$-$(6) and (13)$-$(17) can be solved
self-consistently. Once the solution has been found, the EOS of the
hadronic matter can be calculated from
\begin{eqnarray}
\epsilon=\sum_i\frac{2}{{(2\pi)}^3}\int
d^3k\sqrt{k^2+{m^*_i}^2}[f_i(k)+\overline{f}_i(k)]+\frac{1}{2}m^2_\sigma\sigma^2+\frac{\kappa}{6}g^3_{\sigma
N}\sigma^3\nonumber\\+\frac{\lambda}{24}g^4_{\sigma
N}\sigma^4+\frac{1}{2}m^2_\omega\omega^2+\frac{\xi}{8}g^4_{\omega
N}\omega^4+\frac{1}{2}m^2_\rho\rho^2+3\Lambda_\nu g^2_{\rho N}g^2_{\omega N}\omega^2\rho^2\nonumber\\
+\frac{1}{\pi^2}\sum_l\int
k^2\sqrt{k^2+{m_l}^2}[f_l(k)+\overline{f}_l(k)]dk,
\end{eqnarray}

\begin{eqnarray}
p=\sum_i\frac{1}{3}\frac{2}{{(2\pi)}^3}\int
d^3k\frac{k^2}{\sqrt{k^2+{m^*_i}^2}}[f_i(k)+\overline{f}_i(k)]-\frac{1}{2}m^2_\sigma\sigma^2-\frac{\kappa}{6}g^3_{\sigma
N}\sigma^3\nonumber\\-\frac{\lambda}{24}g^4_{\sigma
N}\sigma^4+\frac{1}{2}m^2_\omega\omega^2+\frac{\xi}{24}g^4_{\omega
N}\omega^4+\frac{1}{2}m^2_\rho\rho^2+\Lambda_\nu g^2_{\rho N}g^2_{\omega N}\omega^2\rho^2\nonumber\\
+\frac{1}{3\pi^2}\sum_l\int
\frac{k^4}{\sqrt{k^2+{m_l}^2}}[f_l(k)+\overline{f}_l(k)]dk.
\end{eqnarray}
Meanwhile, the single-particle energy for the $\Lambda$ hyperons,
and the $\Lambda$ chemical potential at a given baryon density
$\rho_B$ can also be obtained, which are crucial in the
investigation on the superfluidity of $\Lambda$ hyperons.

\section{Energy gap equation for $\Lambda$ hyperon pairing}
In this work, we investigate the $^1S_0$ superfluidity of $\Lambda$
hyperons in warm strange hadronic matter. As a key quantity to determine the onset of superfluidity, the energy gap function
$\Delta_k$ can be obtained by solving the finite-temperature gap
equation
\begin{eqnarray}
\Delta_k=-\frac{1}{\pi}\int
k'^2dk'V(k,k')\frac{\Delta_{k'}}{E_{k'}}\mathrm{tanh}(\frac{E_{k'}}{2T}),\\
E_{k'}=\sqrt{[\varepsilon(k')-\mu_\Lambda]^2+\Delta^2_{k'}},
\end{eqnarray}
where $\varepsilon(k')$ is the single-particle energy in the nuclear
medium for the $\Lambda$ hyperons, $\mu_\Lambda$ the corresponding
chemical potential at a given baryon density $\rho_B$. The
single-particle energy of $\Lambda$ hyperons in the RMF approach is
written as
\begin{eqnarray}
\varepsilon(k)=\sqrt{k^2+{m^*_\Lambda}^2}+g_{\omega\Lambda}\omega
\end{eqnarray}
The potential matrix element for the $^1S_0$ $\Lambda$ pairing
interaction can be given by
\begin{eqnarray}
V(k,k')=\langle k|V_{\Lambda\Lambda}(^1S_0)|k'\rangle=\int r^2
drj_0(kr)V_{\Lambda\Lambda}(r)j_0(k'r),
\end{eqnarray}
where $j_0(kr)=\mathrm{sin}(kr)/(kr)$ is the zero order spherical Bessel function. 
$V_{\Lambda\Lambda}(r)$ is the $^1S_0$ $\Lambda$ pairing interaction potential in coordinate space. 
It is known that the magnitude of the pairing gap are influenced by the
$\Lambda\Lambda$ interaction. Here, we adopt the ND1, ND2, ESC00,
NSC97b, NSC97e and NSC97f potentials obtained by fitting to the
corresponding Nijmegen models \cite{Hiyama,Rijken,Filikhin}, as well
as the NFs and NSC97s potentials obtained by reproducing the
$\Lambda\Lambda$ binding energy value of
$B_{\Lambda\Lambda}(\tiny{^{6}_{\Lambda\Lambda}}$He) from the Nagara
event \cite{Takahashi}. To solve the gap equation, we follow the
separation method developed by Khodel {\it et al.} \cite{Khodel}.

\section{Model parameters}
To examine the influence of the isoscalar-isovector
cross-interaction on the superfluidity of $\Lambda$ hyperons, we
employ the parameter set FSU in the present calculation, meanwhile,
the parameter sets NL3 and NL3* are also included for comparison. As
for the meson-hyperon couplings, the values in the SU(6) quark model are taken
for the vector coupling constants
\begin{eqnarray}
g_{\omega\Lambda}=g_{\omega\Sigma}=2g_{\omega\Xi}=\frac{2}{3}g_{\omega
N},\nonumber\\
g_{\rho \Lambda}=0, ~~g_{\rho\Sigma}=2g_{\rho\Xi}=2g_{\rho N}.
\end{eqnarray}
The scalar coupling constant of $\Lambda$ hyperons is chosen to fit
the $\Lambda$ hypernuclei observables so as to reproduce the
descriptions of $\Lambda$ hypernuclei and $\Lambda$ hyperons in
nuclear matter self-consistently. Studying for the $\Lambda$
hypernuclei with FSU, NL3, and NL3*, in Ref. \citen{xu1}, we
obtained the relative $\sigma$ coupling for $\Lambda$ as
$R_{\sigma\Lambda}$=0.619 for the FSU , $R_{\sigma\Lambda}$=0.620
for the NL3* and $R_{\sigma\Lambda}$=0.621 for the NL3 parameter
set. The $R_{\sigma\Lambda}$ is defined as
$R_{\sigma\Lambda}=g_{\sigma\Lambda}/g_{\sigma N}$. With these
scalar couplings, we obtain the potential depth of a $\Lambda$ in
saturated nuclear matter as: $U^{(N)}_\Lambda=-29.99$ MeV for FSU,
$U^{(N)}_\Lambda=-30.31$ MeV for NL3* and $U^{(N)}_\Lambda=-30.42$
MeV for NL3, which are all close to the reasonable $\Lambda$ hyperon
potential $U^{(N)}_\Lambda\simeq-30$ MeV \cite{Gal}. For
$\Sigma$ hyperons, The study of $\Sigma^-$ atoms showed strong
evidence for a sizable repulsive potential in the nuclear core
\cite{Batty1,Mares}. A recent work again confirmed the repulsive
nature of the $\Sigma^-$ potential with a new geometric analysis of
the $\Sigma^-$ atom data \cite{Friedman}. Therefore, for the
$\Sigma$-$N$ interaction, we adopt $U^{(N)}_\Sigma$= 30 MeV as used
in Ref. \citen{Gal} to determine the sacalar coupling constants.
Besides, for the $\Xi$-$N$ interaction, we take the potential
$U^{(N)}_\Xi= -15$ MeV \cite{Khaustov,Ishizuka} in our calculation.
Then, we obtain, $g_{\sigma\Sigma}=4.717$ and $g_{\sigma\Xi}=3.161$
for the NL3 parameter set, $g_{\sigma\Sigma}=4.650$ and
$g_{\sigma\Xi}=3.119$ for the NL3* parameter set, and
$g_{\sigma\Sigma}=4.820$ and $g_{\sigma\Xi}=3.279$ for the FSU
parameter set, respectively.

For systematically investigate the influence of the
isoscalar-isovector cross-interaction on the superfluidity of
$\Lambda$ hyperons, we change the $\Lambda_\nu$ in our calculation.
For a given $\Lambda_\nu$, we follow Refs. \citen{Horowitz,Jiang}
and \citen{Cavagnoli} to readjust the $\rho NN$ coupling constant
$g_\rho$ so as to keep the symmetry energy unchanged at $k_F= 1.15$
$\mathrm{fm}^{-1}$. This simple procedure produces a nearly constant
binding energy per nucleon for $^{208}$Pb as $\Lambda_\nu$ is
changed \cite{Horowitz}. The readjusted parameters with various
$\Lambda_\nu$ are listed in Table 1, where the parameter sets are
named according to the value of $\Lambda_\nu$, except for the
original parameter sets FSU.
\begin{table}[pt]
\tbl{Readjusted parameters in NL3* and FSU. The binding energy per
nucleon ($E/A$), proton radius ($r_p$), and neutron skin thickness
($r_n-r_p$) for $^{208}$Pb are listed.}
{\begin{tabular}{@{}cccccc@{}} \toprule model&$\Lambda_\nu$&$g_\rho$&$E/A$ (MeV)&$r_p$ (fm)&$r_n-r_p$ (fm)\\
\colrule
FSUw1&0.01&9.6550&-7.866&5.461&0.259\\
FSUw2&0.02&10.5558&-7.871&5.466&0.233\\
FSU&0.03&11.7673&-7.873&5.473&0.206\\
FSUw4&0.04&13.5221&-7.868&5.482&0.175\\
NL3*w1&0.01&9.8047&-7.882&5.452&0.256\\
NL3*w2&0.02&10.6242&-7.896&5.456&0.226\\
NL3*w3&0.03&11.6909&-7.903&5.463&0.196\\
NL3*w4&0.04&13.1605&-7.905&5.472&0.164\\ \botrule
\end{tabular}}
\end{table}

\section{Numerical results and discussion}
\begin{figure}[b]
\centerline{\psfig{file=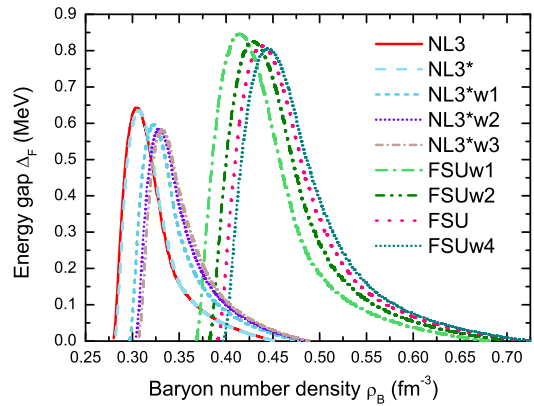}} \caption{Density dependence of
the $^1S_0$ $\Lambda$ energy gap $\Delta_F$ in $\beta$-stable
hadronic matter at $T=0$ with NL3, NL3* and FSU, as well as the
readjusted parameters in NL3* and FSU. The used $\Lambda\Lambda$
interaction is ESC00 potential.}
\end{figure}

In this section, we study the $^1S_0$ $\Lambda$ superfluidity in warm strange hadronic matter.
Firstly, for investigating the influence of the isoscalar-isovector
cross-interaction on the superfluidity of $\Lambda$ hyperons, we
show the energy gap $\Delta_F$ of the $\Lambda$ hyperons in hadronic
matter at $T=0$ with several parameter sets in Fig. 1. Where, the
$\Delta_F$ is the energy gap of $\Lambda$ hyperons at the Fermi
surface. Considering the computing time, here, we just use the ESC00
potential for example. As we see in Fig. 1, The threshold density of
$\Lambda$ is around 0.39 fm$^{-3}$ calculated with FSU. In the case
of FSU,  a $^1S_0$ $\Lambda$ superfluid is formed as soon
as the $\Lambda$ hyperons appear in hadronic matter.
Besides, the same result can also be
obtained for the other eight parameter sets in our calculation. It
is found that the maximal energy gap $\Delta_F$ of $\Lambda$ with
the ESC00 potential for the case of FSU is about 0.81 MeV, while the
result is about 0.63 MeV for both NL3* and NL3, which is in accord
with the result in Ref. \citen{Wang}. Additionally, in Fig. 1, we
find that the maximal energy gap gradually becomes smaller as the
nonlinear coupling $\Lambda_\nu$ increases for both the NL3* and
FSU. Besides, the baryon density corresponding to the maximal energy
gap increases with the nonlinear coupling $\Lambda_\nu$ for NL3* and
FSU, respectively.

\begin{figure}[th]
\centerline{\psfig{file=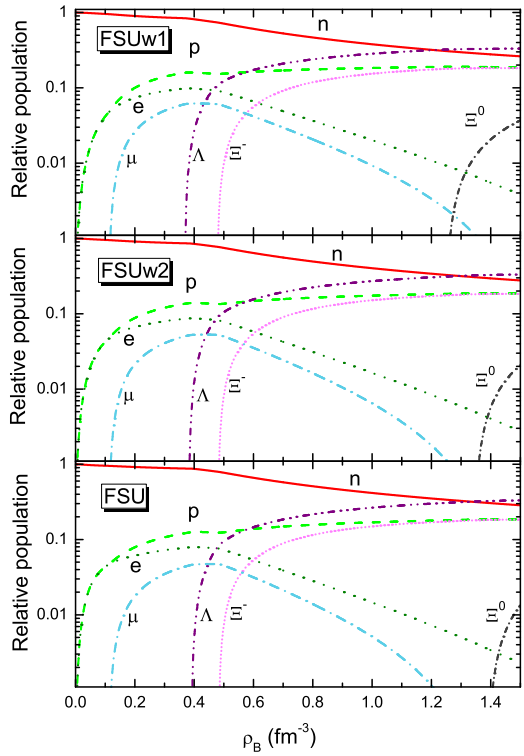}} \caption{Calculated variation
behavior of the relative populations of the compositions of hadronic
matter with respect to the total baryon density $\rho_B$ at $T$=0 in
FSUw1, FSUw2 and FSU, respectively.}
\end{figure}

\begin{figure}[th]
\centerline{\psfig{file=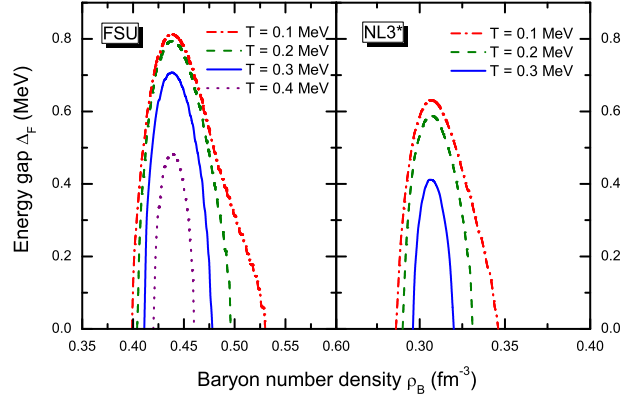}} \caption{Density dependence of
the $^1S_0$ $\Lambda$ energy gap $\Delta_F$ in $\beta$-stable
hadronic matter with FSU at $T$=0.1, 0.2, 0.3 and 0.4 MeV, and with
NL3* at $T$=0.1, 0.2 and 0.3 MeV. The used $\Lambda\Lambda$
interaction is ESC00 potential.}

\end{figure}
\begin{figure}[th]
\centerline{\psfig{file=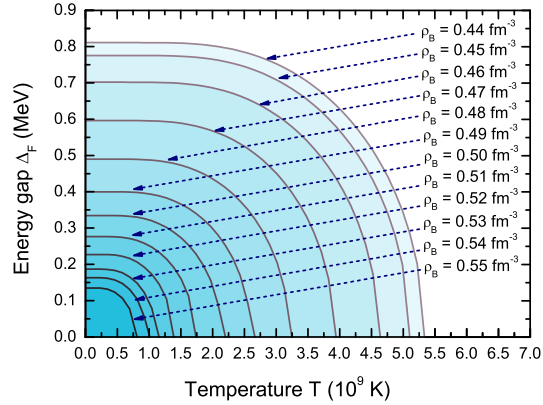}} \caption{Temperature
dependence of the $^1S_0$ $\Lambda$ energy gap $\Delta_F$ in
$\beta$-stable hadronic matter with FSU.  The baryon density
$\rho_B$ ranges from 0.44 fm$^{-3}$ to 0.55 fm$^{-3}$. The used
$\Lambda\Lambda$ interaction is ESC00 potential.}
\end{figure}

\begin{table}[pt]
\tbl{The critical temperature of $^1S_0$ $\Lambda$ superfluid at
different baryon densities. The quantities $Y$ and $\Delta_F$
represent the fraction of $\Lambda$ hyperon ($Y=\rho_\Lambda/\rho_B$
) and the $^1S_0$ $\Lambda$ energy gap in $\beta$-stable hadronic
matter. $T_c^{cal.}$ and $T_c^{WCA}$ are the calculated critical
temperature with ESC00 potential in the case of FSU and the
corresponding weak-coupling approximation (WCA) estimations.}
{\begin{tabular}{@{}ccccc@{}} \toprule $\rho_B$ (fm$^{-3}$)&Y&$\Delta_F$(MeV)&$T_c^{cal.}$($10^9$ K)&$T_c^{WCA}$($10^9$ K)\\
\colrule
0.41&1.25\%&0.502&3.36&3.33\\
0.42&2.14\%&0.697&4.64&4.61\\
0.43&3.08\%&0.791&5.22&5.23\\
0.44&4.06\%&0.810&5.34&5.36\\
0.45&5.06\%&0.775&5.11&5.13\\
0.46&6.06\%&0.702&4.64&4.65\\
0.47&7.06\%&0.597&3.94&3.95\\
0.48&8.05\%&0.490&3.25&3.24\\
0.49&8.85\%&0.400&2.67&2.65\\
0.50&9.50\%&0.335&2.20&2.22\\
0.51&10.10\%&0.271&1.74&1.79\\
0.52&10.67\%&0.223&1.39&1.47\\
0.53&11.22\%&0.182&1.16&1.20\\
0.54&11.75\%&0.163&1.01&1.08\\
0.55&12.26\%&0.135&0.81&0.89\\ \botrule
\end{tabular}}
\end{table}

For the properties of $\Lambda$ hyperons in nuclear medium also play
an important role on the $\Lambda$ energy gap, in Fig. 2, we depict
the relative populations of all compositions with respect to the
total baryon density at $T=0$. Here, we take the results with FSUw1,
FSUw2 and FSU for example. For the $\Lambda$ hyperons, their
relative populations increase rapidly with the ascent of the density
near the onset region of $\Lambda$ hyperons, then they increase very
slowly with the increase of the baryon density at higher density.
The results show that the threshold density of $\Lambda$ increases
with the nonlinear coupling $\Lambda_\nu$ increasing. As shown in
Fig. 2, the cross-interaction may have an important influence on the
properties of $\Xi^0$ hyperons in nuclear medium, which needs more
discussion in future work.

In order to investigate the temperature dependence of the
superfluidity of $\Lambda$ hyperons, in Fig. 3, we show the energy
gap $\Delta_F$ of the $\Lambda$ hyperons in hadronic matter in FSU,
and NL3* included for comparison. For both of the two parameter
sets, with the temperature increasing, the maximal energy gap of
$\Lambda$ pairing decreases, besides, the onset density of $\Lambda$
superfluidity becomes higher and the density where $\Lambda$
superfluidity disappear becomes lower. However, the density
corresponding to the maximal energy gap is almost unchanged with the
temperature increasing, which is about 0.44 fm$^{-3}$ in the case of
FSU and about 0.31 fm$^{-3}$ in NL3*. For better understanding, in
Fig. 4, we show the temperature dependence of the energy gap
$\Delta_F$ of $\Lambda$ for several values of the total baryon
number density in FSU for example. Fig. 4 is obtained with the range
of the temperature $T$ from 0 to 0.48 MeV, with the steps being 0.01
MeV. Finally it should be pointed out that the finite-temperature
gap equation solution is a very lengthy calculation. It takes more
than 600 CPU hours with one processor. As shown in Fig. 4, the
energy gaps of $\Lambda$ at different baryon density are almost
unchanged in low temperature region, then they decrease rapidly with
the increase of the temperature and disappear at some critical
temperature.

\begin{figure}[th]
\centerline{\psfig{file=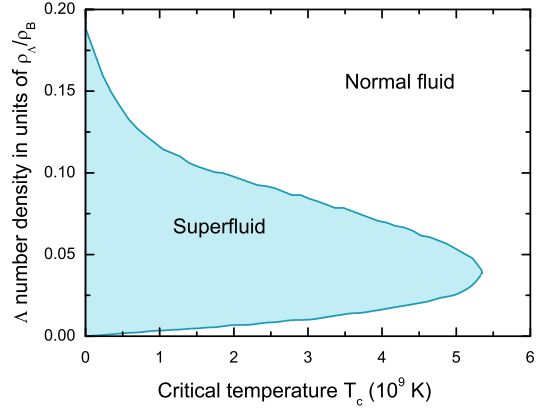}} \caption{Critical temperature
of the $^1S_0$ $\Lambda$ superfluid as a function of the $\Lambda$
number density in units of $\rho_\Lambda/\rho_B$ calculated with FSU
parameter set. The used $\Lambda\Lambda$ interaction is ESC00
potential.}
\end{figure}

\begin{table}[bth]
\tbl{Maximal pairing gap $\Delta_F$ at Fermi surface and the
critical temperature of $^1S_0$ $\Lambda$ superfluid obtained with
several $\Lambda\Lambda$ potentials. $\rho_B$ is the total baryon
density of $\beta$-stable hadronic matter corresponding to the
maximal pairing gap, and $T_c^{cal.}$ is the calculated critical
temperature. The used parameter sets are FSU and NL3*.}
{\begin{tabular}{@{}cccccccc@{}} \toprule
&\multicolumn{3}{c}{FSU}&&\multicolumn{3}{c}{NL3*}\\
\cline{2-4}\cline{6-8}
&$\rho_B$&$\Delta_F$&$T_c^{cal.}$&&$\rho_B$&$\Delta_F$&$T_c^{cal.}$\\
&(fm$^{-3}$)&(MeV)&($10^9$ K)&&(fm$^{-3}$)&(MeV)&($10^9$ K)\\
\colrule
ESC00&0.439&0.81&5.34&&0.306&0.63&4.19\\
ND1&0.432&0.18&1.16&&0.304&0.14&0.92\\
ND2&0.429&0.12&0.73&&0.301&9.8$\times10^{-2}$&0.65\\
NSC97b&0.402&2.0$\times10^{-3}$&1.3$\times10^{-2}$&&0.286&1.7$\times10^{-3}$&1.1$\times10^{-2}$\\
NSC97e&0.406&4.3$\times10^{-3}$&2.7$\times10^{-2}$&&0.289&3.7$\times10^{-3}$&2.3$\times10^{-2}$\\
NSC97f&0.400&1.7$\times10^{-3}$&1.0$\times10^{-2}$&&0.286&1.4$\times10^{-3}$&8.8$\times10^{-3}$\\
NSC97s&0.414&1.1$\times10^{-2}$&6.7$\times10^{-2}$&&0.293&9.9$\times10^{-3}$&5.6$\times10^{-2}$\\
NFs&0.432&1.7$\times10^{-2}$&0.10&&0.304&1.5$\times10^{-2}$&0.09\\
\botrule
\end{tabular}}
\end{table}

In Table 2, we list the $\beta$-stable fractions of the $\Lambda$
and the critical temperature, as well as the estimated result from
the well-known weak-coupling approximation (WCA) \cite{Lifshitz}.
\begin{eqnarray}
T_c\approx0.57 \Delta_F(T=0).
\end{eqnarray}
As seen in Table 2, the calculated critical temperature of the
$^1S_0$ $\Lambda$ superfluid is in good agreement with the WCA. In
our case of FSU, the fractions of the $\Lambda$ in warm hadronic
matter, corresponding to the maximal critical temperature, is about
4\%. Then, with the fractions of the $\Lambda$ increasing, the
corresponding critical temperature decreases. When the fractions of
the $\Lambda$ approaches to 19\%, there will be no $^1S_0$ $\Lambda$
superfluid any more in our calculation. In addition, our calculation
about the maximal energy gap $\Delta_F$ of $\Lambda$ as well as the
corresponding fractions of the $\Lambda$ are in accord with the
result obtained by using the G-matrix effective interaction in Ref.
\citen{Balberg}, where they found that a maximum gap energy of
0.8$-$0.9 MeV is achieved for a $\Lambda$ fraction of about 5\%.

In Fig. 5, we show the region in the temperature-$\Lambda$-density
plane where the $\Lambda$ hyperon is expected to be superfluid. In
the case of FSU, the $\Lambda$ is in a $^1S_0$ superfluid state for
fractions of the $\Lambda$ ranging from $3.8\times10^{-3}$ \% up to
$\sim$18.8\%, which corresponds to a total baryon density ranging
from the $\Lambda$ onset density 0.39 fm$^{-3}$ to $\sim$0.71
fm$^{-3}$ at $T=0$. With the temperature increasing, the density
region for $\Lambda$ superfluid becomes narrow. As seen in Fig. 5,
the $^1S_0$ $\Lambda$ superfluid only exists when the stellar matter
cools down to about 5.4$\times10^9$ K. Above this maximal critical
temperature of $\Lambda$ superfluid, there still may exist $^1S_0$
$\Sigma^-$ pairing and $^3P_2$ neutron pairing \cite{Vidana}.

Finally, we list the maximal pairing gap ($\Delta_F$) at the Fermi
surface and the critical temperature obtained with several
$\Lambda\Lambda$ potentials in Table 3. The total baryon density
corresponding to the maximal pairing gap calculated with FSU is
around 0.4 fm$^{-3}$ with these $\Lambda\Lambda$ potentials, while
it is around 0.3 fm$^{-3}$ in the case of NL3*. However, the maximal
pairing gap $\Delta_F$ and the  critical temperature predicted with
FSU and NL3* are similar to each other. Additionally, the critical
temperature of $^1S_0$ $\Lambda$ superfluid is around $10^7$ K with
the NSC97s and NFs potentials in the case of FSU and NL3*.

\section{Summary}
In this article, the $^1S_0$ superfluidity of $\Lambda$ hyperons in
warm strange hadronic matter, in $\beta$ equilibrium by including
the full octet of baryons, is investigated within the RMF models. By
changing the strength of the isoscalar-isovector cross-interaction
in RMF models (NL3* and FSU), we systematically investigate the
influence of the cross-interaction term on the properties of $^1S_0$
$\Lambda$ superfluid. It is found that with the isoscalar-isovector
coupling increasing, the onset density and the density corresponding
to the maximal energy gap of $^1S_0$ $\Lambda$ superfluid increases.
However, the maximal energy gap of $\Lambda$ pairings gradually
becomes smaller with the isoscalar-isovector coupling. In addition,
it is found that the maximal energy gap $\Delta_F$ of $\Lambda$ in
$\beta$-stable hadronic matter at $T$= 0 is about 0.81 (0.63) MeV
with the ESC00 potential in the case of FSU (NL3*). The value of
$\Delta_F$ is $0.1-0.2$ MeV for the ND1 and ND2 potentials. The
NSC97b, NSC97e, NSC97f, NSC97s and NFs potentials reproduce the
value of $\Delta_F$ of the order of $10^{-3}-10^{-2}$ MeV.

On the other hand, with the temperature increasing, the onset
density of $^1S_0$ $\Lambda$ superfluid becomes higher and the
disappearance density becomes lower, while the density corresponding
to the maximal energy gap is almost unchanged. The energy gaps of
$\Lambda$ pairing at different baryon density are almost unchanged
in low temperature region, then they decrease rapidly with the
increase of the temperature and disappear at some critical
temperature. The maximal critical temperature of $\Lambda$
superfluid is about 5.3 (4.2)$\times10^9$ K in the case of FSU
(NL3*) with the ESC00 potential. The maximal critical temperature of
$\Lambda$ superfluid is of the order of $10^{7}-10^{9}$ K with the
ND1, ND2, NSC97b, NSC97e, NSC97f, NSC97s and NFs potentials.

\section*{Acknowledgements}
The authors would like to thank the anonymous referee for her/his
constructive suggestions which are very helpful to improve this
manuscript. This work was supported by the National Natural Science
Foundation of China (Grants No. 11035001, No. 11375086, No.
10735010, No. 10975072, No. 11120101005, No. 11175232 and No.
11105072), by the 973 National Major State Basic Research and
Development of China (Grants No. 2013CB834400, No. 2014CB845402 and
No. 2010CB327803), by CAS Knowledge Innovation Project No.
KJCX2-SW-N02, by Research Fund of Doctoral Point (Grant No.
20100091110028), by the Project Funded by the Priority Academic Program Development of
Jiangsu Higher Education Institutions (PAPD), and by the Research and Innovation Project for College
Postgraduate of JiangSu Province, Grants No. KYZZ\_0023.

\end{document}